\documentclass{ws-ijmpd}

\usepackage[super,compress]{cite}
\usepackage{graphicx}
\usepackage{epstopdf}
\usepackage{amsfonts}
\usepackage{amssymb}
\usepackage{epsfig}
\usepackage{hyperref}

\def\de#1/de#2{\frac{\partial {#1}}{\partial {#2}}}

\newcommand{\ba}{\begin{eqnarray}}
\newcommand{\ea}{\end{eqnarray}}
\newcommand{\be}{\begin{equation}}
\newcommand{\ee}{\end{equation}}

\begin{document}


%
\catchline{}{}{}{}{}
%

\title{Correlation of structure growth index with current cosmic acceleration: constraints on dark energy models
}

\author{G.~Panotopoulos}

\address{
Centro de Astrof{\'i}sica e Gravita{\c c}{\~a}o-CENTRA, Instituto Superior T{\'e}cnico-IST, Universidade de Lisboa-UL, Av. Rovisco Pais, 1049-001 Lisboa, Portugal. \\
Departamento de Ciencias F{\'i}sicas, Universidad de la Frontera, Casilla 54-D, 4811186 Temuco, Chile.
\\
\href{mailto:grigorios.panotopoulos@ufrontera.cl}{\nolinkurl{grigorios.panotopoulos@ufrontera.cl}} 
}

\author{G.~Barnert}

\address{
Departamento de Astronom{\'i}a, FCFM, Universidad de Chile, Camino El Observatorio 1515, Las Condes, 
Santiago, Chile.
\\
\href{mailto:gerald_barnert@hotmail.com}{\nolinkurl{gerald\_barnert@hotmail.com}} 
}

\author{L.~E.~Campusano}

\address{
Departamento de Astronom{\'i}a, FCFM, Universidad de Chile, Camino El Observatorio 1515, Las Condes, 
Santiago, Chile.
\\
\href{mailto:luis@das.uchile.cl}{\nolinkurl{luis@das.uchile.cl}} 
}

\maketitle

\begin{history}
\received{Day Month Year}
\revised{Day Month Year}
\end{history}

\begin{abstract}
We study dynamical dark energy models within Einstein's theory by means of matter perturbations and the growth 
index $\gamma$. Within four-dimensional General Relativity, we assume that dark energy does not cluster, and we 
adopt a linear ansatz for the growth index to investigate its impact on the deceleration parameter, $q$, and on the dark energy equation-of-state parameter, $w$. Following this approach, we identify a relationship between $q_0$ (today's value of $q$) and $\gamma$, which to the best of our knowledge is new. For $w(z)$, we find that in most of the cases considered it crosses the -1 line (quintom) ending at a present day value $w_0 > -1$. Furthermore, we show that an analytic expression for  $w(z)$ may be obtained in the form of order (4,4) (or higher) Pad{\'e} parameterizations. 
\end{abstract}


\ccode{}

\section{Introduction}\label{Intro}

Brightness measurements of supernovae over a wide redshift range 25 years ago surprisingly revealed consistency with an accelerating Universe \cite{SN1,SN2}. Current cosmic acceleration can be simply explained through a new fluid component of the Universe dubbed Dark energy (DE) \cite{turner}, whose origin and nature present a pressing challenge to current theoretical Cosmology. Einstein's General Relativity (GR) \cite{GR} framework applied to a Universe containing only radiation and non-relativistic matter provide cosmological equations whose solutions do not allow for accelerating solutions. Instead, such a behavior of the Universe is allowed when a positive cosmological constant \cite{einstein,carroll} is introduced into Einstein's equations, which together with collisionless dark matter constitute the basis of the 
$\Lambda$CDM model.  

\smallskip

The standard cosmological model ($\Lambda$CDM model) is consistent with a broad range of observational tests, however since some time now it has been facing two discrepancies, one related to the cosmological constant problem \cite{zeldovich,weinberg,Sahni:1999gb}, and the other concerning the value of the Hubble constant, $H_0$. On the latter, the value of $H_0$ needed to fit the CMB anisotropy measurements and the one calculated from the distances and recession velocities of low red-shift galaxies, differ significantly, see e.g. \cite{tension,tension1,tension2,tension3}—dubbed "Hubble tension". The value of the Hubble constant found by both the PLANCK Collaboration \cite{planck1,planck2} and the Atacama Cosmology Telescope (ACT) \cite{act}, $H_0 = (67-68)~\text{km/(Mpc sec)}$, is about $10 \%$ lower than the value determined using local galaxies, $H_0 = (73-74)~\text{km/(Mpc sec)}$ \cite{hubble,recent,recent2}. If the discrepancy between the two measurements of $H_0$ turns out to be real, then it might be indicative of the need of new physics \cite{newphysics,earlyDE,ben}. For recent reviews on challenges for the $\Lambda$CDM model see \cite{Perivolaropoulos:2021jda,Peebles2022}.

\smallskip

An independent determination of the Hubble constant, \cite{Abbott} might be possible using the second order Taylor expansion for the luminosity distance as a function of red-shift, $d_L$ \cite{hubble,Ref2}
\begin{equation}
d_L(z) \approx \frac{z}{H_0} \: \left[ 1+\frac{z}{2} (1-q_0) \right].
\end{equation}
This provides us with an allowed range for the deceleration parameter, $q_0$, \cite{Ref2} derived from a $q_0-j$ parameterization, where $j$ is the jerk parameter (or one of the two statefinders). 
The determination of the Hubble constant through simultaneous observations of red-shift and luminosity distance can be achieved by studying inspiraling neutron star binaries, where the absolute distance of the source can be determined by gravitational waves measurements \cite{sirens}, while short gamma-ray bursts signatures in the electromagnetic band may allow the determination of red-shift \cite{Ref1}.

\smallskip

Several dark energy models have been proposed as possible alternatives to the $\Lambda$CDM model in an attempt to solve, 
or at least to alleviate, the cosmological constant problem and the tension associated to the Hubble constant. Those models may be, broadly speaking, classified into two categories. First, modified gravity models, where additional terms and/or curvature corrections are added to GR at cosmological scales, e.g. $f(R)$ theories of gravity \cite{mod1,mod2,HS,starobinsky}, brane-world models \cite{langlois,maartens,dgp}, and Scalar-Tensor theories of gravity \cite{BD1,BD2,Boisseau:2000pr,leandros,PR}. Second, dynamical equation of state (EoM) parameter $w<1/3$ models, e.g, quintessence \cite{DE1}, phantom \cite{DE2,DE3}, quintom \cite{DE4,DE5}, tachyonic \cite{DE6} or k-essence \cite{DE7}. For an excellent review on the dynamics of dark energy see e.g. \cite{copeland}. 

\smallskip

Depending on the details of the underlying theory of gravity and/or the properties of the assumed DE model, the evolution
of linear matter perturbations may be affected in several ways. Even if two DE models predict the same late-time accelerating expansion of the Universe, they still may differ in how matter perturbations \cite{staro,radouane}evolve and
grow with time. This fact could provide an additional important way to discriminate at low red-shift between various DE models, see e.g. \cite{Boisseau:2000pr,extra1,extra2,extra3,extra4,Calderon:2019jem,Calderon:2019vog}. It is therefore important to characterize as accurately as possible the growth of matter perturbations. In particular, a quantity that has been extensively studied over the years in the literature is the so called growth index, $\gamma$, introduced by Wang and Steinhardt in Ref. \cite{gamma}. As a matter of fact, the growth rate of matter perturbations may be probed by means of three-dimensional weak lensing surveys \cite{Verde}.

\smallskip

In this work we study in detail the growth of matter perturbations, and give a thorough discussion on the correlation between a linear ansatz for the growth index with both the deceleration parameter and the DE equation-of-state parameter. Contrary to the common approach, in which one first adopts a concrete DE model and then solves the equation for the matter density contrast to compute the growth index, here we follow the inverse approach. Namely, first we assume that the growth index is not a constant (namely characterized by a non-vanishing derivative), and then investigate its impact on properties of dark energy, integrating numerically the equation for matter perturbations assuming a homogeneous (i.e. it does not cluster) dark energy. What is more, we use here a relatively new constraint on the deceleration parameter (to be defined in the next section), $q_0 = -0.50 \pm 0.08$, coming from standard sirens to put bounds on the growth index, see the discussion in section 3 after eq. (22).

\smallskip

The present article is organized as follows: In the next section we briefly review the spatially flat 
Friedmann-Robertson-Walker Universe at background level as well as the basics of linear cosmological perturbation theory. 
In section 3 we show and discuss our main numerical results. Finally, in section 4 we summarize our work with some concluding remarks. We adopt the mostly positive metric signature, $(-,+,+,+)$, and the speed of light in vacuum is set to unity, $c=1$.

\section{Theoretical framework}

\subsection{Background evolution}

The set of cosmological equations governing the expansion history of a spatially flat, homogeneous and isotropic Universe may be found e.g. in \cite{review}. Although they are standard textbook expressions, to set the notation and for self-completeness, we collect here the definitions and the equations we will be using for the numerical analysis to be 
presented below in section 3.

\smallskip

On the gravity side, a spatially flat, isotropic and homogeneous Universe is described by a Robertson-Walker metric \cite{review}
\begin{equation}
ds^2 = -dt^2 + a(t)^2 \delta_{ij} dx^i dx^j,
\end{equation}
where $t$ is the cosmic time, while $a(t)$ is the scale factor. On the matter side, a perfect fluid with total pressure 
$p$ and total energy density $\rho$, is characterized by an energy-momentum tensor given by \cite{review}
\begin{equation}
T_{\nu}^{\mu} =\textrm{diag}(-\rho,p,p,p)
\end{equation}

\smallskip

Within Einstein's theory the cosmological equations are found to be the two Friedmann equations as well as the continuity 
equation \cite{review}
\begin{eqnarray}
H^2 & = & \frac{8 \pi G}{3} \rho \\
\frac{\ddot{a}}{a} & = & -\frac{4 \pi G}{3} (\rho + 3p) \\
0 & = & \dot{\rho} + 3 H (\rho+p),
\end{eqnarray}
where an over dot denotes differentiation with respect to $t$, and $H \equiv \dot{a}/a$ by definition is the Hubble parameter. 

\smallskip

Next, the deceleration parameter, $q$, is defined by
\begin{equation}
q \equiv - \frac{\ddot{a}}{aH^2} = -1 + (1+z) \frac{H'(z)}{H(z)},
\end{equation}
where $z=-1+a_0/a$ is the red-shift, and with $a_0=1$ being the present value of the scale factor. 

\smallskip

Finally, introducing for convenience dimensionless quantities, $E \equiv H/H_0$ and $\Omega_m \equiv \rho_m / \rho_c$, with
$H_0$ being the Hubble constant, and where $\rho_c \equiv \frac{3H^2}{8 \pi G}$ is the critical density, 
the first Friedmann equation takes the form
\begin{eqnarray}
E(a)^2 & = & \Omega_{m,0} a^{-3} + (1 - \Omega_{m,0}) F(a) \\
w(z) & = & -1 + (1+z) \: \frac{F'(z)}{3 F(z)}
\end{eqnarray}
neglecting radiation at late times, where $\Omega_{m,0}$ is today's matter normalized density, while $w(z)$ is the DE equation-of-state parameter as a function of the red-shift. In the following, the independent variable will be taken to be either the scale factor or the red-shift, instead of the cosmic time.

\subsection{Linear cosmological perturbations}

Now we briefly review linear cosmological perturbation theory within General Relativity, see e.g. \cite{Mukhanov:2005sc,other}.

\smallskip

The main goal is to solve the perturbed Einstein's field equations
\begin{equation}
\delta G_\nu^\mu = 8 \pi G \: \delta T_\nu^\mu.
\end{equation}
For scalar perturbations, relevant to the growth of structures, the metric tensor takes the form
\begin{equation}
ds^2 = -(1 + 2 \Psi) dt^2 + (1-2 \Psi) \delta_{ij} dx^i dx^j,
\end{equation}
where the potential $\Psi$ is the metric perturbation. Regarding the cosmological fluid, the perturbed 
stress-energy tensor takes the form
\begin{equation}
\delta T_0^0 = - \delta \rho, \qquad \delta T_j^i = \delta p \: \delta_j^i.
\end{equation}
The full set of coupled equations for the perturbations of the system "metric tensor plus perfect fluid" 
may be found e.g. in \cite{mariam,pano1,pano2}.

\smallskip

Any perturbation $g(t,\vec{x})$ depends both on time and spatial coordinates, and it satisfies a linear partial differential equation. However, after performing a Fourier transform, the density contrast, $\delta_k = \delta \rho_m / \rho_m$, with $k$ being the wave number, for pressure-less matter satisfies the following linear ordinary differential equation \cite{rogerio,leandros2,review2}
\begin{equation}
\ddot{\delta_k} + 2H \dot{\delta_k} - 4 \pi G \rho_m \delta_k = 0
\end{equation}
within linear perturbation theory, $\delta_k \ll 1$, for sub-horizon scales, $k/(2 \pi a) \gg a H$, assuming that only non-relativistic matter clusters. It is not difficult to verify that during matter domination
\begin{equation}
a(t) \sim t^{2/3}, \; \; \; \; \; H(t) = \frac{2}{3 t}
\end{equation}
the matter density contrast grows linearly with the scale factor, $\delta_k(a) \sim a$.

\smallskip

If now the scale factor is taken to be the independent variable, the equation for $\delta$ may be take equivalently the following form 
\begin{align}
\delta''(a) + \left( \frac{3}{a} + \frac{E'(a)}{E(a)} \right) \delta'(a) - \frac{3}{2} \frac{\Omega_m}{a^5 E(a)^2} \delta(a) = 0,
\end{align}
where for simplicity we drop the sub-index $k$, and a prime now denotes differentiation with respect to the scale factor.

\smallskip

Finally, the growth index, $\gamma$, is defined by
\begin{eqnarray}
f & \equiv & \frac{d \ln \delta}{d \ln a} = \frac{a}{\delta} \: \frac{d \delta}{d a} \\
f & = & \Omega_m^\gamma.
\end{eqnarray}
During the epoch of matter domination, since in the matter era $\delta \propto a$, it is not difficult to verify
that $f(a)=1$. Furthermore, the equation for $\delta$ may be written down equivalently as a new equation for $f(a)$ as follows
\begin{equation}
a f'(a) + f(a)^2 + f(a) \left( 2 + a \frac{E'(a)}{E(a)} \right) = \frac{3 \Omega_m(a)}{2}.
\end{equation}

\section{Numerical results}

In the simplest case in which the DE equation-of-state is a constant, $w(z)=w$, it turns out that there is an impressive agreement between the numerical result and an analytic approximation, and at lowest order it is computed to be \cite{leandros2}
\begin{equation}
\gamma = \frac{3(w-1)}{6w-5}
\end{equation}
around $z \sim 1$. This reduces to $\gamma=6/11$ for the special case corresponding to the $\Lambda$CDM model (for which $w=-1$). In full generality, however, the growth index evolves with time, and it is a function of red-shift, with a non-vanishing derivative $\gamma'(z) \equiv d\gamma(z)/dz$. In an attempt to further improve on the analytic approximation, in \cite{radouane,leandros2} the authors considered an expansion to first order in $z$ of the form
\begin{equation} \label{linear}
\gamma(z) = \gamma_0 + \gamma_1 z,
\end{equation}
which could have interesting observational consequences, and which is characterized by two parameters. Those may be 
identified with the present values of the functions $\gamma(z),\gamma'(z)$
\begin{equation}
\gamma_0 = \gamma(z=0), \;\;\;\;\;  \gamma_1 = \gamma'(z=0).
\end{equation}

As already mentioned, in the present work instead of adopting a specific DE model, i.e. a given parameterization
$w(a)$ or $w(z)$, we assume, following \cite{radouane,leandros2}, that the evolution of matter perturbations implies a linear growth index on the red-shift space of the form eq. (\ref{linear}) valid at low red-shift in the range 
$0 \leq z \leq 1$. The parameters $\gamma_0,\gamma_1$ typically vary in 
the range \cite{radouane}
\begin{equation}
0.5 < \gamma_0 < 0.6, \; \; \; \; \; -0.05 < \gamma_1 < 0.05,
\end{equation}
and we investigate what the impact of those two free parameters is on properties of DE, such as the equation-of-state parameter, $w(z)$, and on the deceleration parameter evaluated at present, $q_0$. In contrast to \cite{radouane}, where only an algebraic equation relating $w_0,\gamma_0,\gamma_1$ was considered, we integrate numerically the equation for matter perturbations, and determine how both $q$ and $w$ evolve with red-shift. Moreover, we put bounds on $\gamma$ using an allowed range on $q_0$ that stems from a $(q_0-j)$ parameterization \cite{Ref2}, with $j$ being the jerk parameter.


\begin{figure}
\centering
\includegraphics[width=0.7\linewidth]{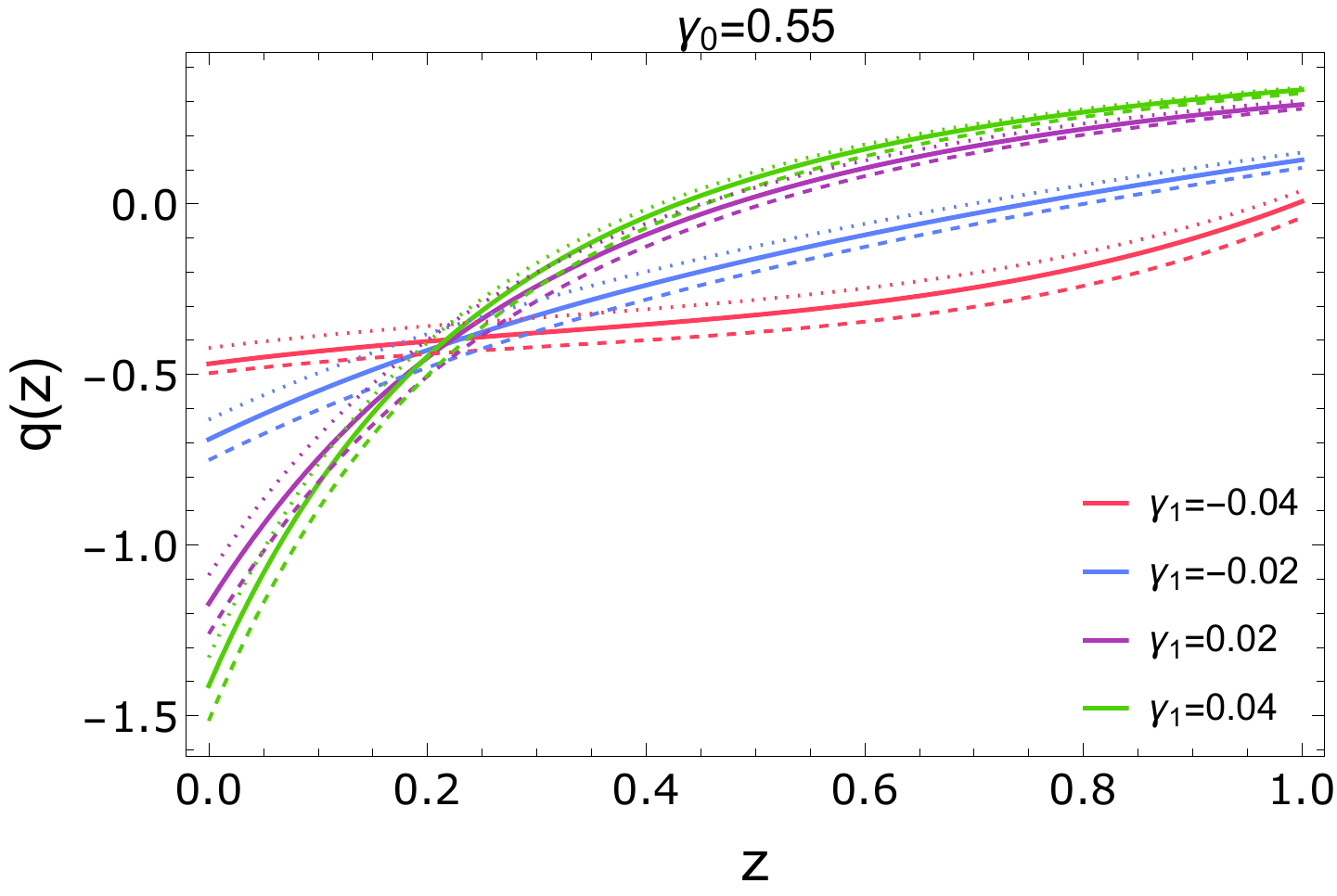} \
\includegraphics[width=0.7\linewidth]{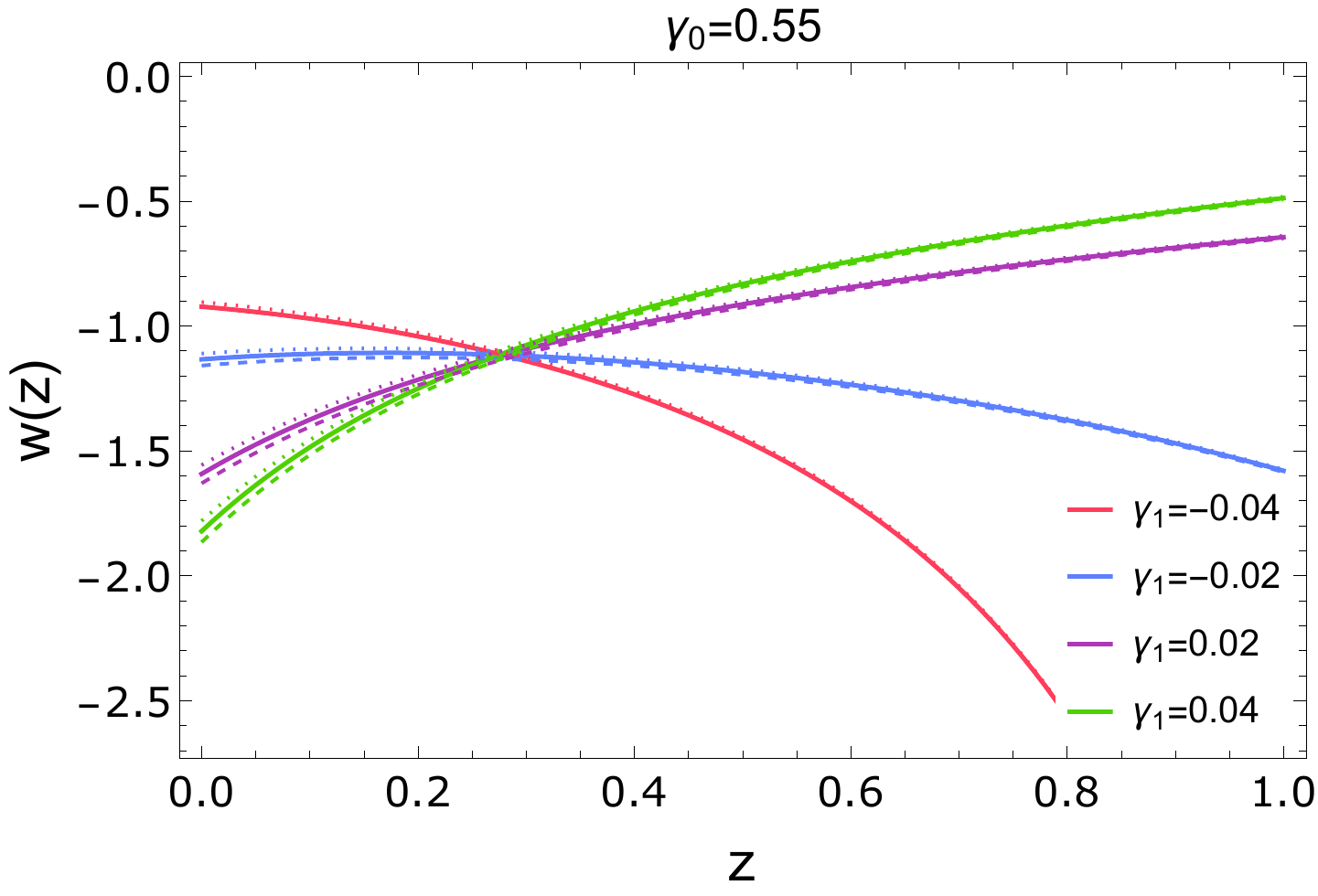} \
\caption{
{\bf TOP:} Deceleration parameter, $q$, versus red-shift, $z$, for $\Omega_{m,0}=0.28,0.30,0.32, \gamma_0=0.55$ and 
$\gamma_1=-0.04, -0.02, 0.02, 0.04$.
{\bf BOTTOM:} Equation-of-state parameter, $w$, versus red-shift, $z$, for $\Omega_{m,0}=0.28, 0.30, 0.32, \gamma_0=0.55$ and 
$\gamma_1=-0.04, -0.02, 0.02, 0.04$.
}
\label{fig:2}
\end{figure}


\begin{figure}
\centering
\includegraphics[width=0.7\linewidth]{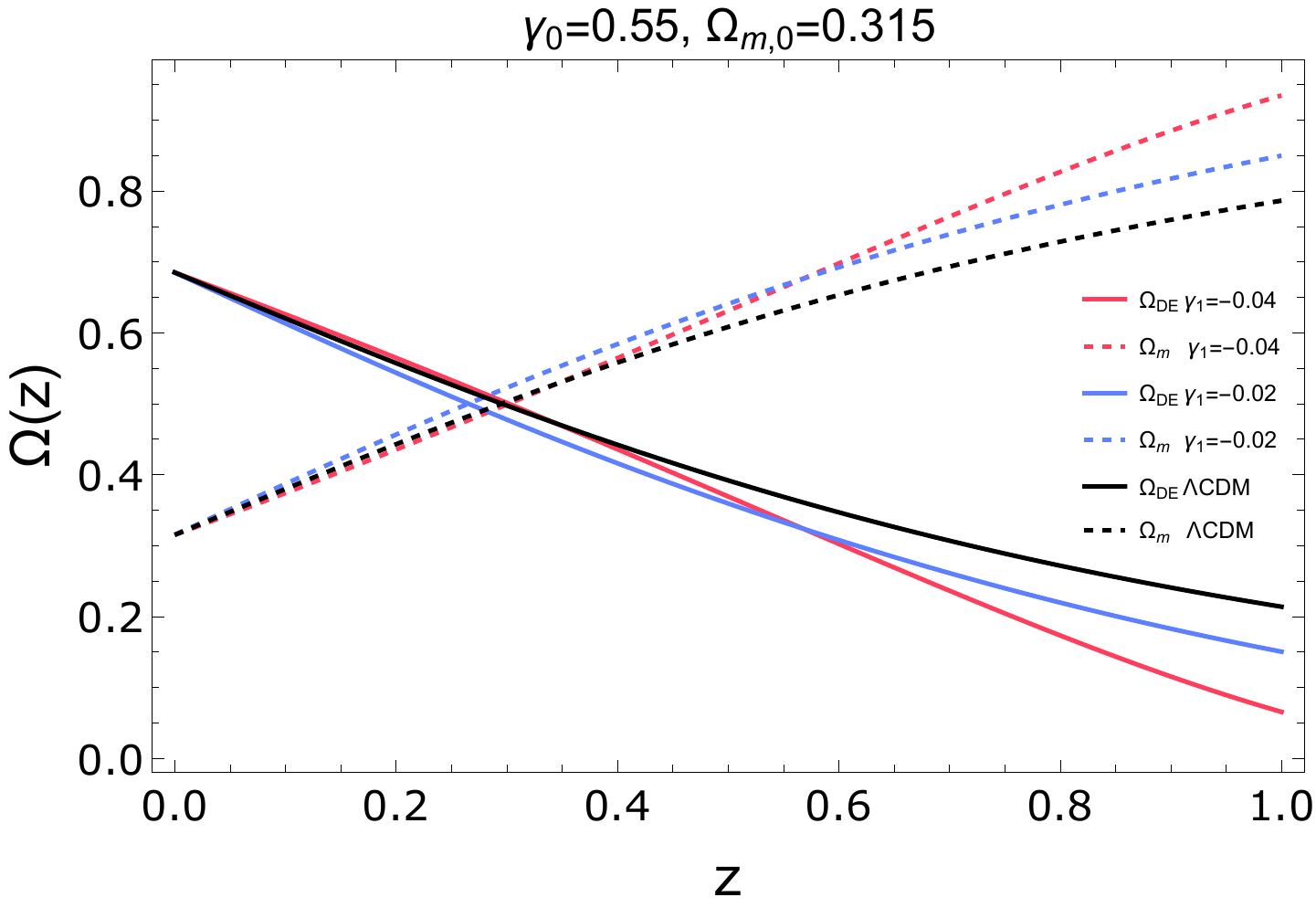} 
\caption{
Comparison between dynamical DE studied here and best fit $\Lambda$CDM setting $\Omega_{m,0}=0.315$ \cite{planck2}: Normalized densities, $\Omega_m,\Omega_{DE}=1-\Omega_m$, versus red-shift, $z$, varying $\gamma_1$ and setting 
$\gamma_0=0.55$. 
}
\label{fig:3}
\end{figure}


\begin{figure*}[ht]
\centering
\includegraphics[width=0.45\textwidth]{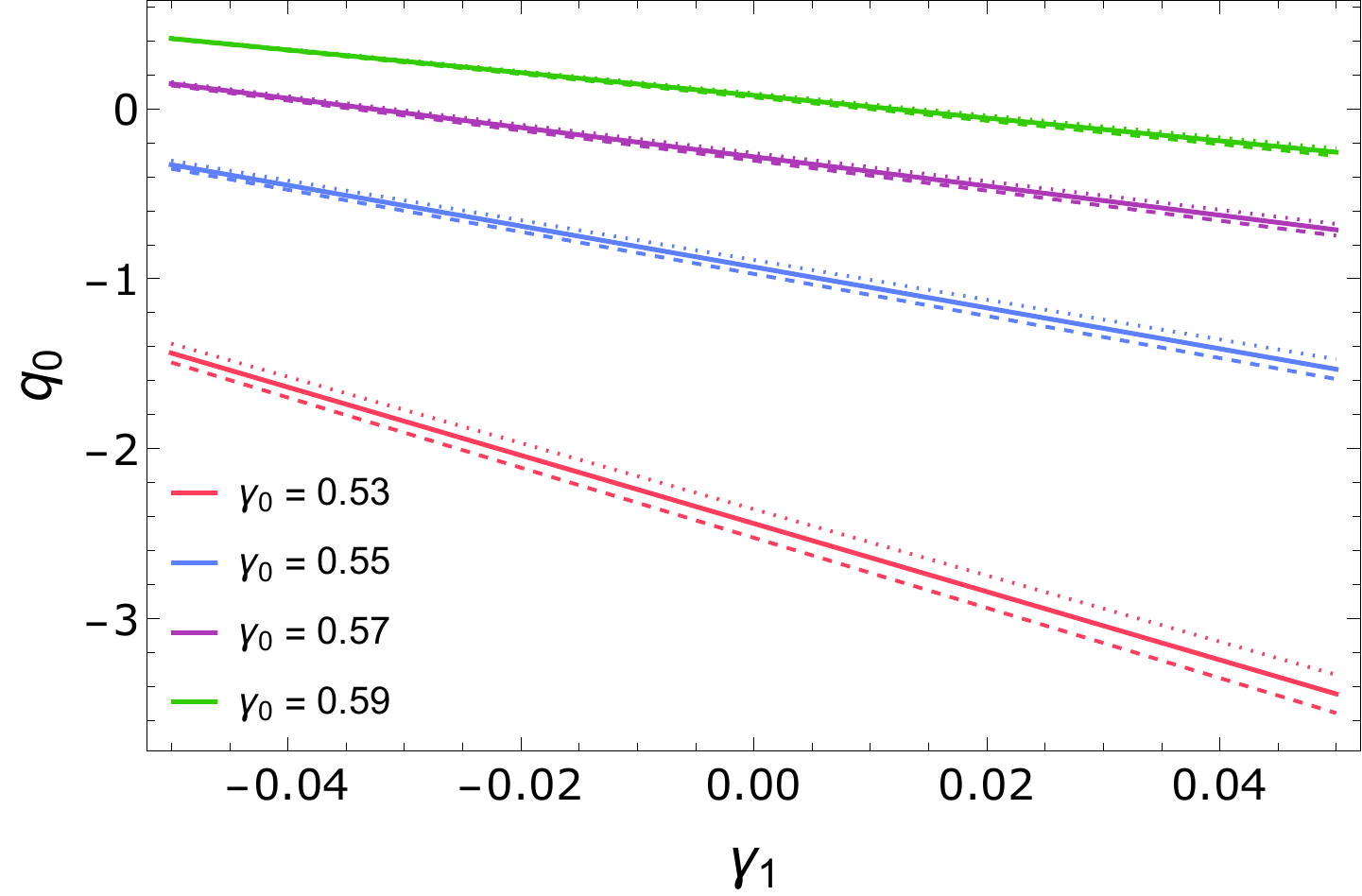}   \
\includegraphics[width=0.47\textwidth]{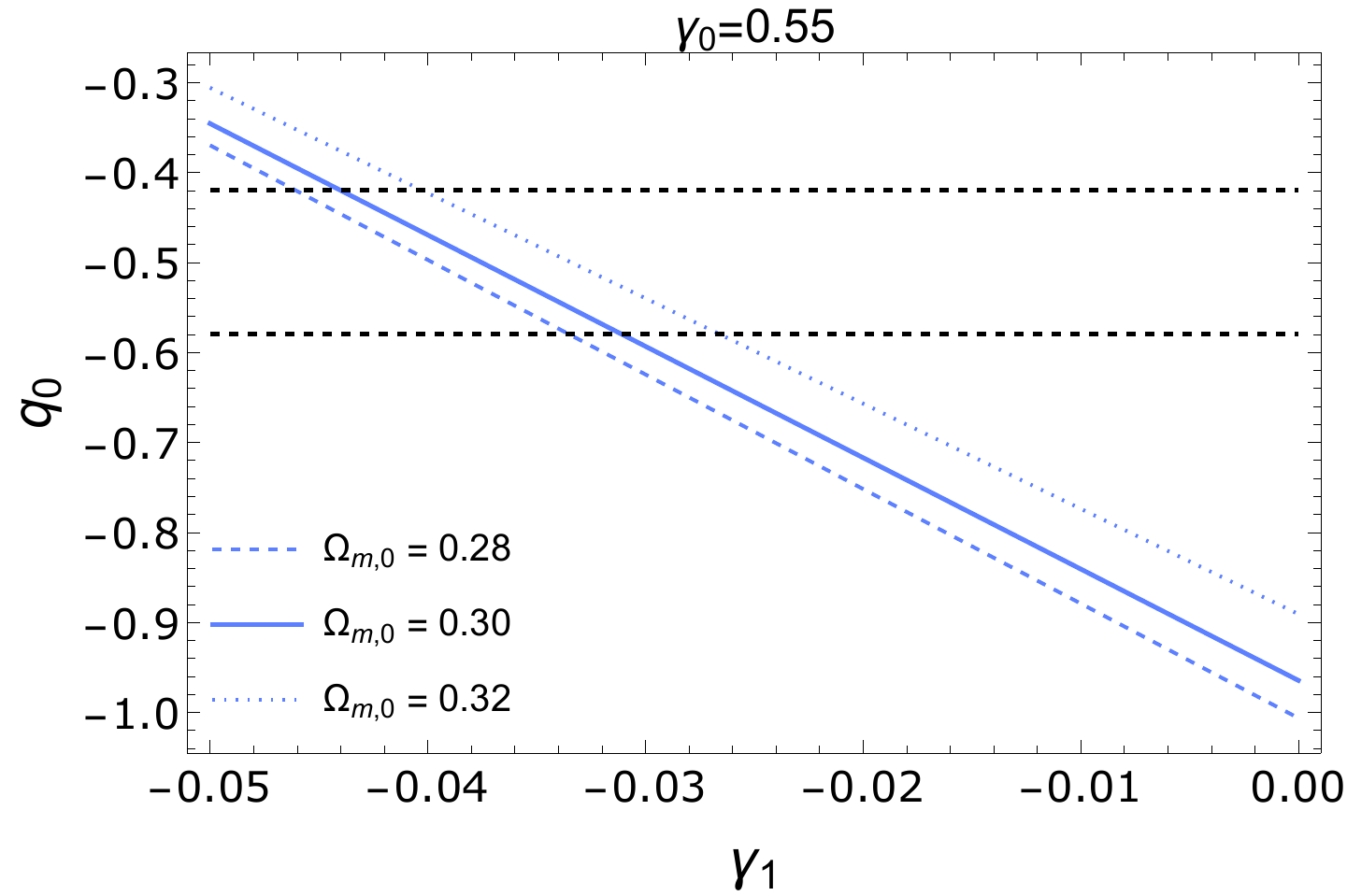}   \
\includegraphics[width=0.45\textwidth]{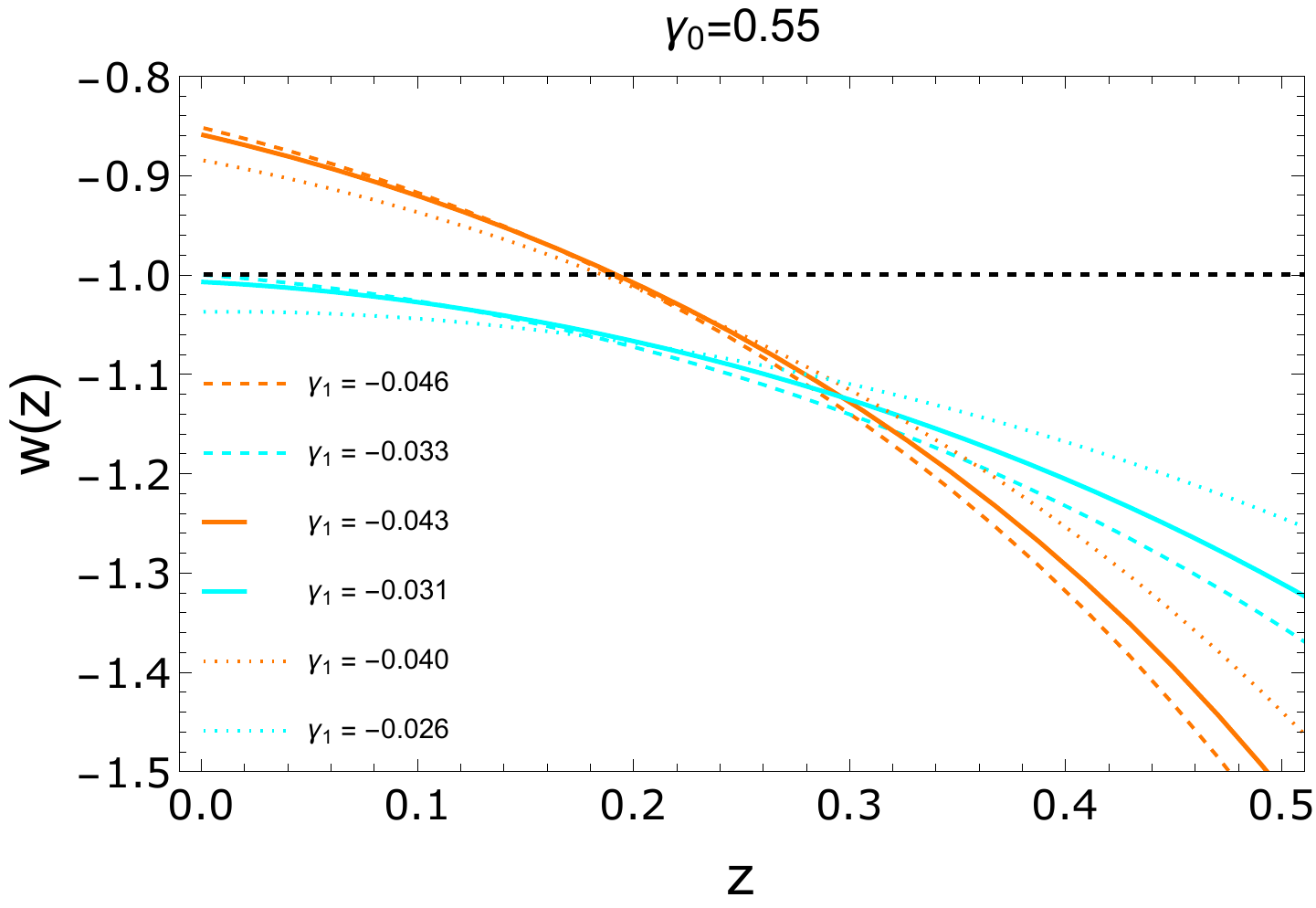}   \
\caption{
{\bf Top left:} Deceleration parameter evaluated at today, $q_0$, versus $\gamma_1$ for $\Omega_{m,0}=0.28, 0.30, 0.32$ 
and $\gamma_0=0.53,0.55,0.57,0.59$.
{\bf Top right:} $q_0$ versus $\gamma_1$ for $\Omega_{m,0}=0.28, 0.30, 0.32$ and $\gamma_0=0.55$. The limits $q_0=-0.50 \pm 0.08$ are shown as well.
{\bf Bottom:} Equation-of-state parameter, $w$, versus red-shift, $z$, for $\Omega_{m,0}=0.28, 0.30, 0.32$ and 
$\gamma_0=0.55$.
}
\label{fig:1}
\end{figure*}


\smallskip

Let us briefly describe the approach followed here. Since the DE model is a priori unknown, all quantities of interest may be expressed in terms of the unknown function $F(a)$, which is directly related to the DE equation-of-state $w(a)$. The differential equation for $f(a)$ may be viewed as a differential equation for $F(a)$ instead, subjected to the initial condition $F(a=1)=1$. Once the numerical values of 
$\Omega_{m,0},\gamma_0,\gamma_1$ are specified, the differential equation may be integrated numerically, and after that $q(z)$ and $w(z)$ may be computed in a straightforward manner. 

\smallskip

Our main numerical results are displayed in the Figures below. Throughout the numerical analysis we vary both $\gamma_0$ 
and $\gamma_1$ as well as the fractional density of matter considering $\Omega_{m,0}=0.28 \; (\textrm{dashed curves}), 0.30 \; (\textrm{solid curves}), 0.32 \; (\textrm{dotted curves})$. 

\smallskip

First, the impact of the $\gamma_1$ variation on $q(z)$ (upper panel) and $w(z)$ (lower panel) is shown in Fig.~\ref{fig:2} setting $\gamma_0=0.55$ and varying $\Omega_{m,0}=0.28, 0.30, 0.32$. We have considered four different values of 
$\gamma_1=-0.04,-0.02,0.02,0.04$. The equation-of-state parameter crosses the $-1$ line (quintom) either from lower to higher values when $\gamma_1$ is negative or from higher to lower values when $\gamma_1$ is positive. Moreover, both $q_0$
and the transition red-shift, $z_*$, between deceleration and acceleration decrease with $\gamma_1$. It is worth noticing that in both panels of Fig.~\ref{fig:2}, all four curves meet (approximately) at the same point. To be more precise, after a closer look we have observed that regarding $q(z)$, the curves meet at $z \approx 0.22$, while regarding $w(z)$ the curves meet at $z \approx 0.28$. Although different, the two distinct points lie one close to another, and in fact 
they correspond roughly to the point at which $\Omega_m = \Omega_{DE}$ (see Fig.~\ref{fig:3}), or in other words when DE starts to dominate the expansion of the universe. From that point of view, it does not come as a surprise, although at the moment we do not have a deeper explanation to offer as to why this happens.

\smallskip

Next, in the left panel of Fig.~\ref{fig:1} we show $q_0$ versus $\gamma_1$ for a certain value of $\gamma_0$. The four different curves correspond to distinct numerical values of $\gamma_0=0.53, 0.55, 0.57, 0.59$ from bottom to top. For a given $\gamma_0$, $q_0$ decreases almost linearly with $\gamma_1$, while as $\gamma_0$ increases the curves are displaced upwards, and they are less rapidly varying functions of $\gamma_1$. 

\smallskip

Similarly, in the right panel of Fig.~\ref{fig:1} we show $q_0$ versus $\gamma_1$ for a fixed 
$\gamma_0=0.55$, and include the horizontal strip corresponding to the allowed range $q_0 = -0.50 \pm 0.08$. 
Thus, we obtain for $\gamma_1$ the allowed range depending on $\Omega_m$ as follows
\begin{equation}
-0.046 \leq \gamma_1 \leq -0.033, \; \; \; \; \Omega_m=0.28,
\end{equation}
\begin{equation}
-0.043 \leq \gamma_1 \leq -0.031, \; \; \; \; \Omega_m=0.30,
\end{equation}
\begin{equation}
-0.040 \leq \gamma_1 \leq -0.026, \; \; \; \; \Omega_m=0.32.
\end{equation}

For the two extreme values corresponding to the lower and upper bound of $\gamma_1$, we show in the bottom panel of Fig.~\ref{fig:1} the equation-of-state parameter $(z)$. In the case represented by the cyan curve, $w(z)$ remains always below the $-1$ line, and therefore DE is phantom \cite{DE2,DE3}. At a more fundamental level based on a Lagrangian description, this class of dynamical DE model may be analyzed introducing a real scalar field with the "wrong" sign in front of its kinetic term. On the other hand, the case represented by the orange curve crosses the $-1$ line, and therefore DE is quintom \cite{DE4,DE5}. This time, again at a more fundamental level based on a Lagrangian description, this class of dynamical DE model may be analyzed introducing two real scalar fields, one minimally coupled and another with the "wrong" sign in front of its kinetic term \cite{saridakis1,saridakis2}.

\smallskip

Finally, a comment is in order at this stage. Within our approach the DE equation-of-state may be obtained numerically. However, as always it is both desirable and advantageous to have an analytic expression for $w(z)$, since one may study
the behaviour of the function in certain limiting cases for instance. One way to do that is to fit the function $F(z)$ with a polynomial of degree $n$, $F(z) \approx P_n(z)$. Recall that $F(z)$ is directly related to $w(z)$. Then it is easy to verify that the DE equation-of-state parameter takes the form of a ($m,n$) Pad{\'e} parameterization \cite{PADE}, which only recently is discussed in the literature concerning cosmology and DE (see e.g. \cite{pade1,pade2}), and which quite generically for any function $h(z)$ is given by \cite{Spyros1,Spyros2}
\begin{equation}
h(z) = \frac{R_n(z)}{Q_m(z)} = \frac{c_0+c_1 z + ... + c_n z^n}{d_0+d_1 z + ... + d_m z^m},
\end{equation}
namely a rational function where both the numerator and the denominator are polynomials of degree $n$ and $m$, respectively.

\smallskip

To evaluate the goodness of the polynomial fits, namely to check if the fit to a polynomial is a good one, we compute the relative errors with respect to the numerical solution, and we require that it does not exceed $5 \%$ in order for the fit to be a satisfactory one. Although this choice is somewhat arbitrary, we consider it to be informative, while at the same time it is not too restrictive.

\smallskip

We have carefully checked that low degree polynomials are not acceptable according to such a criterion, and show that the Pad{\'e} approximation for the DE EoS parameter must be at least (4,4), in contrast to the analysis performed in a previous related work \cite{pade1}, where a (1,1) Pad{\'e} parameterization was considered.
The main differences between that work and ours are the following. First, the ansatz for $\gamma(z)$ in Ref. \cite{pade1} was not a linear one. Moreover, here we integrate the equation for matter perturbations numerically, whereas in Ref. \cite{pade1} an analytical approach, developed in Ref. \cite{Piazza}, was followed. 

\smallskip

For instance, in the case where $\Omega_{m,0}=0.3, \gamma_0=0.55$ and $\gamma_1=-0.029$, we find the following analytic expression for $w(z)$
\begin{eqnarray}
w(z) & = & \frac{R_4(z)}{Q_4(z)}, \\
R_4(z) & = & 0.33 z^4+1.33 z^3-1.98 z^2-0.03 z-3.25, \\
Q_4(z) & = & z^4-2.08 z^3-0.29 z^2-0.25 z+3.17,
\end{eqnarray}
while in the case where $\Omega_{m,0}=0.3, \gamma_0=0.55$ and $\gamma_1=-0.042$, we find the following analytic expression for $w(z)$
\begin{eqnarray}
w(z) & = & \frac{R_5(z)}{Q_5(z)}, \\
R_5(z) & = & 0.65 z^5 + 1.34 z^4-1.29 z^3 +0.09 z^2-1.37 z-1.32, \\
Q_5(z) & = & z^5 -0.97 z^4-0.4 z^3-1.46 z^2 + 0.59 z + 1.52.
\end{eqnarray}

\section{Conclusions}

In summary, the present work has been devoted to the study of dynamical DE models within four-dimensional GR. Assuming that DE does not cluster, the starting point was the differential equation for the density contrast of non-relativistic matter. Instead of the usual treatment, where one adopts a specific DE model to compute the growth index, here we followed the inverse approach. To be more precise, we first assumed for $\gamma(z)$ a linear ansatz $\gamma(z) = \gamma_0 + \gamma_1 z$,
already considered previously in \cite{radouane,leandros2}, valid at low red-shift, $0 < z < 1$, and then investigated its impact on the properties of DE. Within this approach (where the only assumptions are i) a linear ansatz for the growth index, and ii) a dynamical DE model in which DE does not cluster) we have found a relationship between $q_0$ and $\gamma$, which to the best of our knowledge had not been obtained before. Our main numerical results reveal that for $\Omega_{m,0}, \gamma_0$ fixed, $q_0$ is a decreasing function of $\gamma_1$, whereas for a given $\gamma_1$, the curves are displaced upwards as $\gamma_0$ increases. Furthermore, using the relatively new constraint on $q_0$, $q_0=-0.5 \pm 0.08$, coming from standard sirens, we obtained the allowed range for $\gamma_1$ for a given set of parameters $(\gamma_0=0.55,\Omega_{m,0}=0.28, 0.30, 0.32)$. In addition, for viable models we found that DE is either phantom or $w(z)$ crosses the -1 line from initially very negative values (phantom), ending with a present day value $w_0 > -1$ (quintessence). This behavior, called quintom in the literature, may be obtained at the level of a Lagrangian description introducing two real scalar fields, one of which enters with the wrong sign in front of the kinetic term. Finally, we showed that an analytic expression for the DE equation-of-state parameter may be obtained in the form of Pad{\'e} DE parameterizations, and that the order of the Pad{\'e} approximation must be at least (4,4).


\section*{Acknowlegements}

The author G.~P. thanks the Funda\c c\~ao para a Ci\^encia e Tecnologia (FCT), Portugal, 
for the financial support to the Center for Astrophysics and Gravitation-CENTRA, Instituto 
Superior T\'ecnico, Universidade de Lisboa, through the Grants No. UID/FIS/00099/2020 and 
No. PTDC/FIS-AST/28920/2017. 
L.~E.~C. acknowledges partial support from the Center of Excellence in Astrophysics and Associated 
Technologies (PFB06, CONICYT).


\end{document}